# Engineered Octave Frequency Comb in Integrated Chalcogenide Dual-ring Microresonators

**Zifu Wang[1,2†] Liyang Luo[1,2†], Di Xia[1,2†], Siqi Lu[1,2], Guosheng Lin[1,2], Shecheng Gao[3], Bin Zhang[1,2*], Zhaohui Li[1,2,4]**

[1] Guangdong Provincial Key Laboratory of Optoelectronic Information Processing Chips and Systems, School of Electrical and Information Technology, Sun Yat-sen University, Guangzhou 510275, China.
[2] Key Laboratory of Optoelectronic Materials and Technologies, Sun Yat-sen University, Guangzhou 510275, China.
[3] Department of Electronic Engineering, College of Information Science and Technology, Jinan University, Guangzhou 510275, China.
[4] Southern Marine Science and Engineering Guangdong Laboratory (Zhuhai), Zhuhai, 519000, China.
**\* Correspondence:**
Bin Zhang
zhangbin5@mail.sysu.edu.cn
[†] these authors are equal contributions to this work

## Abstract

Broadband Kerr combs with a flat comb spectral profile are expected in a number of applications, such as high-capacity optical communication. Here, we propose novel concentric dual-ring microresonators (DRMs) for advanced dispersion engineering to tailor the comb spectral profile. The dispersion can be flexibly engineered not only by the cross-section of the DRMs, but also by the gap between concentric dual-ring microresonators, which provides a new path to geometrically control the spectral profile of the soliton Kerr combs. An octave-spanning dissipative Kerr soliton comb with superior spectral flatness has been achieved numerically, covering from the telecommunication band to the mid-infrared (MIR) band region with a -40 dB bandwidth of 1265 nm (99.82 THz). Our results are promising to fully understand the nonlinear dynamics in hybrid modes in DRMs, which helps control broadband comb formation.

## 1    Introduction

Microresonator-based Kerr combs (microcombs) have attracted significant research interest in the past decades, which enables the generation of mode-locked laser pulse in chip-scale photonic devices at milliwatt-level power[1]. Dissipative Kerr solitons (DKSs) in microresonators have been demonstrated for high-quality laser sources with high coherence and large bandwidth, originating from the double balance between the dispersion and nonlinearity as well as the cavity losses and the parametric gain[2]. Up-to-date, soliton microcombs have revolutionized various applications, including large-capacity optical communications[3,4], precision metrology, and molecular spectroscopy[5-7], massively parallel LiDAR[8], chip-scale frequency synthesizer[9], etc[10,11]. Specifically, octave-spanning soliton microcombs enable the high signal-to-noise ratio via phase locking of carrier-envelope-offset frequency ($f_{ceo}$), which becomes a significant task[12].

Bright soliton generation in integrated microresonators, benefitted from cavity-enhanced nonlinear efficiency and lithographically controlled accurate dispersion engineering, has advantages in extending ultra-broadband soliton microcombs [13,14]. With the aid of dispersive waves (DWs), octave-spanning microcombs have been demonstrated in many nonlinear photonic platforms[15-20]. However, the large dispersion barrier between the pump and the locations of DWs inevitably results in a decrease in the flatness of the broadband soliton microcombs. Many attempts at advanced dispersion engineering have been proposed, including slot waveguides[21-24], bilayer structures[25], and multi-cladding schemes[26-28]. Recently, multiple zero group velocity wavelengths have been proven to facilitate flat and broadband microcomb generation, which are theoretically achieved in the integrated microresonators with the above-mentioned methods. However, these microresonators with complex structures are sensitive to fabrication tolerance and remain challenging in high-quality ($Q$) factor photonic systems. Recently, the concentric dual-ring microresonators (DRMs)[29-32] are proposed to potentially manipulate the group velocity dispersion by controlling the mode coupling between the hybrid waveguides, providing additional degrees of freedom for geometric design in comparison to single-ring microresonators (SRMs). However, tailoring the bandwidth and flatness of the soliton microcombs in this attractive structure, has not been explored yet.

In this work, broadband and ultra-flat optical frequency comb generation in DRMs is theoretically investigated based on the photonic-integrated chalcogenide glasses (ChGs) platform[33-37]. The characteristics of supermode coupling and integrated dispersion of DRMs are studied to achieve local anomalous dispersion by mode hybridization between the inner and outer microresonator in DRMs. Accordingly, multiple DWs can be attained by introducing local anomalous dispersion in the strong normal dispersion regime in mid-infrared (MIR), leading to beyond-octave frequency comb generation spanning from 1227 nm to 2912 nm with a -40 dB bandwidth of 1265 nm. The spectral shape of the soliton microcomb can be flexibly tuned by controlling the outer ring width. Moreover, the comb power at a specific spectral region in concentric DRMs is enhanced, which is potential for a broad range of applications, such as coherent optical communications. Our results provide a novel route to achieve broadband-integrated microcombs with a user-defined target spectral profile.

## 2 Operation principle of supermode hybridization in DRMs

The structure of the DRM is shown in Figure 1A, which is composed of two concentrically placed microring resonators with a certain distance and an independent coupling bus waveguide. The new home-developed chalcogenide glass-Ge$_{25}$Sb$_{10}$S$_{65}$ is chosen as the core, and the air upper cladding is used to reduce material absorption in MIR, see Figure 1B. The system is driven by a continuous-wave laser for broadband frequency comb generation.

Local dispersion profiles can be engineered by adequately designing the structural parameters of DRM and introducing mode hybridization. Here, the fundamental quasi-transverse magnetic mode (TM$_{00}$) mode is taken into consideration. In the absence of the coupling, the optical path lengths (OPLs) of the inner and the outer SRM can be described as the following[31],

$$\text{OPL}_{in} = 2\pi R_{in} n_{in} \ , \tag{1}$$

$$\text{OPL}_{\text{out}} = 2\pi R_{\text{out}} n_{\text{out}} \, , \tag{2}$$

where $\text{OPL}_{\text{in}}$ and $\text{OPL}_{\text{out}}$ represent the OPLs of the inner and outer microresonator, respectively. $R_{\text{in}}$ and $R_{\text{out}}$ denote their ring radius. The $n_{\text{in}}$ and $n_{\text{out}}$ are the effective mode refractive indices. The mode coupling between hybrid waveguides occurs at the regime where OPLs of the individual inner and outer SRMs are equal[31].

We perform the modified dispersion simulation of the DRM based on the structural parameters of R = 30 μm, $W_1$ = 2200 nm, H = 650 nm, gap = 850 nm, and $W_2$ = 1050 nm. A cross point of the calculated OPLs of the $TM_{00}$ modes for two separate SRMs appears at the wavelength of around 2.4 μm, see Figure 1C. To characterize the formation of mode hybridization, the free spectral ranges and group velocity dispersion of the DRM are investigated. The resonance frequencies of microresonators are determined by[16],

$$\omega_\mu = \omega_{\text{s}} + D_1\mu + \frac{1}{2}D_2\mu^2 + \cdots \, , \tag{3}$$

Where $D_1/2\pi$ is equivalent to the free spectral range of microresonators, $D_k(k \geq 2)$ is the k-order dispersion coefficient, $\omega_{s(u)}$ is the angular resonant frequency for pump mode and other modes, and $\mu$ is the relative mode number. The integrated dispersion $D_{\text{int}}$ including the full-order dispersion term, can be calculated by[16,17],

$$D_{\text{int}} = \omega_\mu - \omega_{\text{s}} - D_1\mu = \sum_{k=2}^{\infty} \frac{1}{k!} D_k\mu^k \, , \tag{4}$$

The integrated dispersion is related to the dispersive wave generation and flatness of microcomb, which will be discussed in parts 3 and 4. Here, we investigate the impact of mode hybridization on second-order dispersion ($D_2$). In the DRM system, the modes in the inner and outer microresonator will divert to each other as wavelength increases and generate the resonant mode hybridization between the coupled waveguides. Compared with the individual microresonators, the mode distributions in DRMs show the superposition of the original modes in inner and outer microresonators (inset of Figure 1C), which are defined as symmetric and antisymmetric supermode. As the mode hybridization modifies the effective index of supermodes, their eigenfrequencies shift slightly from the original uncoupled modes in SRMs accordingly[29]. Hence, local FSRs change remarkably, and the crosstalk of FSRs arises to modify second-order dispersion of supermodes, as depicted in Figure 1D. The symmetric supermode features strong normal dispersion ($D_2 < 0$) around the mode coupling region, while the antisymmetric supermode undergoes a period of strong anomalous dispersion ($D_2 > 0$), see Figure 1E. Therefore, the antisymmetric mode of the DRMs can introduce anomalous dispersion in the strong normal dispersion region and be utilized to modify the integrated dispersion $D_{\text{int}}$ profile of the microresonators.

## 3    Versatile dispersion engineering

The local dispersion at the specific wavelength of the DRMs can be adjusted by tailoring the gap between the inner microresonator and the outer microresonator (gap) as well as the width of the outer microresonator ($W_2$). Here considering the antisymmetric mode in the same DRM in part 2, increasing the outer ring width $W_2$ or the gap will push the coupling region to a longer wavelength, see Figures 2A and 2C. It is worth noting that the dispersion profile far from the coupling wavelength remains unchanged. Four zero-integrated dispersion wavelengths with a flat spectral shape can be observed by optimizing the geometric parameters of the microresonators, see Figure 2B and 2D. When the structure of the inner ring is fixed, the local anomalous dispersion and integrated dispersion with a tunable coupling position for antisymmetric mode are favorable by adjusting the gap and $W_2$.

## 4    Ultra-flat and broadband frequency comb generation

Generally, higher-order dispersion plays a significant role in the generation of DWs, which can tune the spectral shape[38,39]. Here, we investigate the influence of dispersion engineering of DRMs on the frequency comb generation, especially on multiple DWs generation.

A phase-matching condition between the soliton pulse and DW is required for the generation of DWs, which is defined as,

$$\beta(\omega_d) = \beta(\omega_s) + \frac{(\omega_d - \omega_s)}{v_g} + \frac{\gamma P}{2} , \qquad (5)$$

where $\beta$ is the propagation constant of light, $\omega_d$ and $\omega_s$ are the angular frequencies of DW and pump light, $v_g$ denotes the group velocity, P is pump power, $\gamma$ is the nonlinear coefficient. The phase mismatching induced by nonlinear phase shift $\gamma P/2$, can be negligible under low pump power. Therefore, the spectral position ($\omega_d$) of the DW generation is approximately given by linear phase-mismatching conditions in the fiber system, as defined in equation (5) without the third term on the right-hand side[40]. In chip-based microresonator systems, this linear phase-matching condition is analogous to the integrated dispersion $D_{int}(\mu_{DW}) = 0$ in equation (4), where $\mu_{DW}$ is the relative mode number of DWs[41].

Then, we numerically simulate the spectral and time dynamic of Kerr frequency comb in concentric DRMs and SRMs by mean-field LLE model[42-44],

$$\frac{\partial A}{\partial t} = -\left(\frac{\kappa}{2} - i\Delta\right)A + \sum_{k\geq 2}(-i)^{k+1}\frac{D_k}{k!}\left(\frac{\partial}{\partial\theta}\right)^k A + ig_0|A|^2A + \sqrt{k_c}S_{in}, \qquad (6)$$

where $|A|^2$ is the intracavity photon number, $t = nT_R$ is the slow time with $T_R$ representing the roundtrip time and n representing the number of $T_R$ in simulation. $\theta = \frac{2\pi\tau}{T_R}$ is the angular location of the light field envelope in microresonators with $\tau$ representing the fast time of light. $\Delta = \omega_p - \omega_0$ is the frequency detuning between the pump and the cold cavity resonance. Generally, the pump frequency scan from blue detuning ($\Delta > 0$) to red detuning ($\Delta < 0$) for frequency comb excitation, where the nonlinear thermal phase shift is not included in our model because it makes little difference to the bandwidth of frequency comb[45]. $g_0 = \frac{\hbar\omega^2 c n_2}{n_g^2 V_{eff}}$ is the Kerr gain coefficient related to the nonlinear refractive index $n_2$ and effective modal volume $V_{eff}$. The cavity total decay rate $\kappa = \kappa_0 + \kappa_c = \frac{\omega}{Q_0} + \frac{\omega}{Q_c}$ is composed of two parts, the intrinsic decay rate $\kappa_0$ and external coupling rate $\kappa_c$, which can be derived from the quality factor $Q_0$ and $Q_c$, respectively. In this work, the pump frequency linearly detunes from $5\kappa$ to $-45\kappa$ to trigger the formation of microcomb, and the simulation parameters used in our model are listed in Table 1. We first analyzed the integrated dispersion in a traditional microresonator with R = 30 um, $W_1$ = 1600 nm, and H = 650 nm, see Figure 3A. The maximum phase mismatching between the pump and DW reaches ~ 70 GHz, resulting in strong power depression of the spectral comb lines, which hinders the access of octave frequency comb with high flatness. Therefore, when DW moves farther from the pump wavelength, the spectral region between the pump wavelength and DW will show a larger phase mismatching, causing more energy discrepancy among the comb lines[39,46,47]. For comparison, in DRM with R = 30 um, $W_1$ = 2200 nm, H = 650 nm, $W_2$ = 1050 nm, gap = 825 nm, a flat integrated dispersion curve can be realized with $D_2/2\pi$ = 57.77 MHz, and $D_3/2\pi$ = 1.84 MHz by creating an additional anomalous dispersion region in the long wavelength, see the upper panel in Figure 3B. Three zero integrated dispersion points (excluding pump wavelength) are observed in the wavelength range from 2000 nm to 3000 nm, allowing an overall flat dispersion spectral shape and a smaller phase mismatching (~ 20 GHz) compared with SRM.

In numerical simulation, mode-locked octave frequency comb with a -40 dB bandwidth of 1265.84 nm (99.82 THz) and 620 GHz comb lines spacing can be obtained in DRM when the pump detuning is swept to $-32\kappa$, see the lower panel in Figure 3B. Because of the high Kerr nonlinearity of chalcogenide material, the driving pump power for such broadband soliton microcomb is 40 mW. Compared to that of the SR-MR in Figure 3A, the soliton microcomb of the DR-MR features wide bandwidth with a flat envelope at the same detuning. The dynamic behavior of DWs spectral evolution can be observed as the pump laser detunes from $5\kappa$ to $-45\kappa$. At small detuning $-10\kappa$, three DWs emerge at the spectral position corresponding to $D_{int}$ = 0, as shown in the upper panel in Figure 3B. As the pump wavelength moves further into red detuning, two gradually merge into one with higher comb line power (Figure 3D) due to the power-dependent nonlinear phase shift[48]. Taking the detuning effect into account, we can confirm that the spectral position of DWs is determined by $D_{int}(\mu) = \Delta$[48]. In the temporal domain, the soliton pulse is characterized by two kinds of DW tails (DW1 and DW2) sitting on the background of continuous waves [48], see Figure 3D. The peculiar structures of spectral and temporal profiles in the DRM improve the understanding of broadband soliton comb with multiple DWs. It also highlights the utility of DRMs as a feasible scheme to extend spectral region deep into the MIR footprint region for molecular spectroscopy.

Moreover, the DWs position can be flexibly tuned while keeping the large bandwidth in the DRMs to meet the high demands of practical applications. For example, the integrated dispersion curves at near 2500 nm are engineered to adjust the number and the position of the zero integrated dispersion points by increasing the width of the outer microresonator of the DRMs, see Figure 4A., broadband (octave) DKS spectra accompanied by the generation of DW can be observed correspondingly when the width of the outer microresonator is 1060nm. Moreover, as $W_2$ decreases, the positions of the DWs tend to shift to longer wavelengths, giving rise to beyond-octave DKS, see Figure 4B. Therefore, devisable DKS states can be obtained by tailoring the spectral profile of DWs, allowing for the extension of spectral coverage and boosting comb outpower at the desired wavelengths[38].

## 5    Conclusion

In summary, we have systematically investigated the effect of advanced dispersion engineering of the DRMs on the spectral evolution of soliton microcombs generation. By introducing the mode hybridization, the dispersion can be spectrally optimized in favor of generating multiple dispersive waves. Octave-spanning Kerr combs with the target shape can be realized numerically by geometrically controlling the DRMs. This flexible DRMs structure enables mode coupling in hybrid waveguides and control of the spectral location of the dispersive wave, which is critical in broadband soliton microcombs generation.

## Data Availability Statement

## Funding

This work was supported by Key Project in Broadband Communication and New Network of the Ministry of Science and Technology (MOST) (2018YFB1801003), the National Key R&D Program of China under Grant (2019YFA0706301), National Science Foundation of China (NSFC) (U2001601, 61975242, 61525502, 11974234), the Science Foundation of Guangzhou City (202002030103).

## Conflict of interest

The authors declare no conflict of interest.

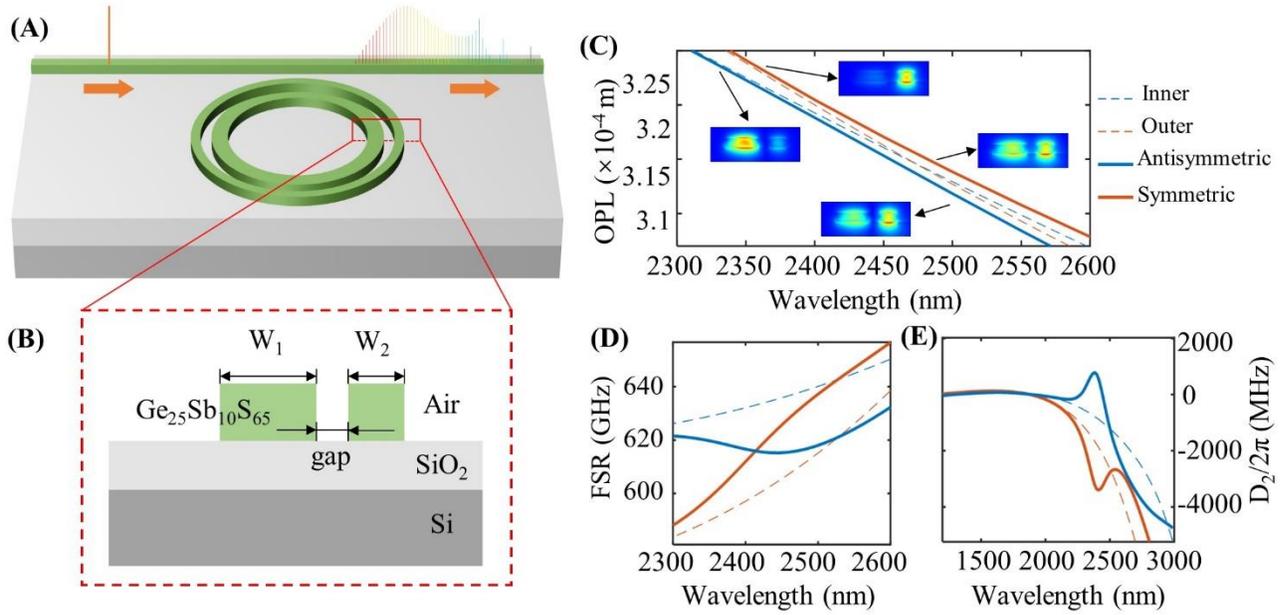

FIGURE 1. Schematic of the proposed DRM and characterization of supermode hybridization. The DRM has an inner ring of radius R = 30 μm, an inner ring width of $W_1$ = 2200 nm, a height of H = 650 nm, a gap of 850 nm, and an outer ring of $W_2$ = 1050 nm. (A) 3D profile of the DRM. (B) Cross section of the DRM based on $Ge_{25}Sb_{10}S_{65}$ material on insulator. (C) The variation of optical path lengths (OPLs) and corresponding electric mode field distributions. The dashed lines are the calculated OPLs of the $TM_{00}$ modes in the inner and outer SRM, respectively. The solid lines are OPLs of antisymmetric and symmetric supermodes in the DRM. (D) The calculated free spectral ranges (FSRs) and (E) group velocity dispersion ($D_2$) of the fundamental transverse magnetic modes ($TM_{00}$).

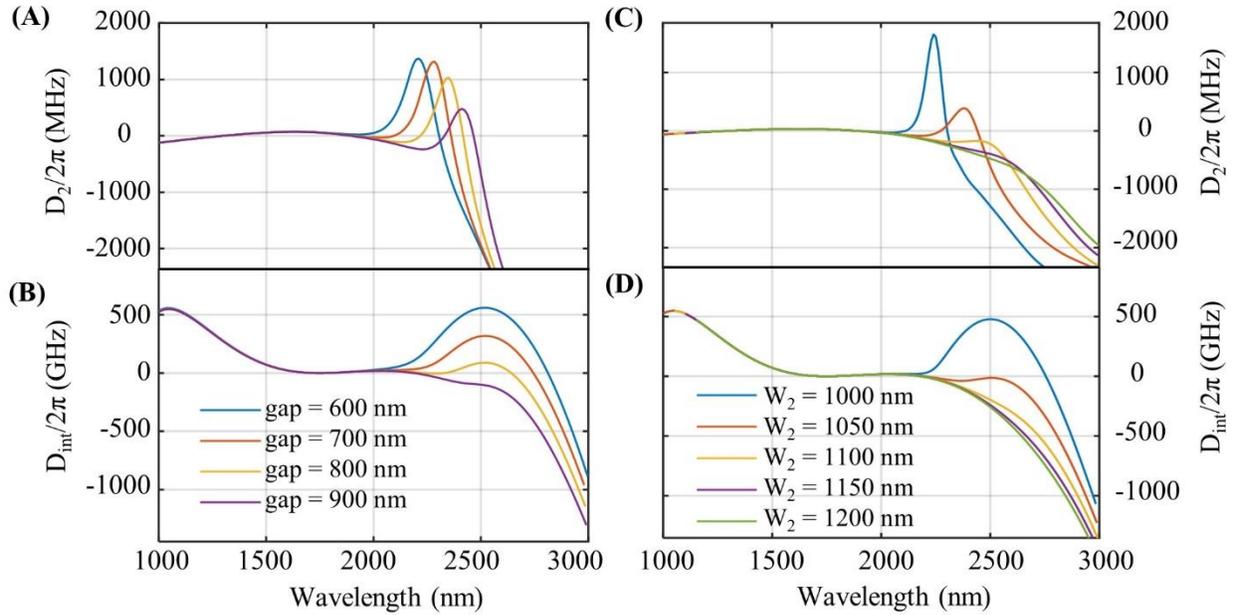

FIGURE 2. Dispersion profiles of the DRMs with structural parameters varying at near R = 30 μm, $W_1$ = 2200 nm, H = 650 nm, gap = 850 nm, $W_2$ = 1050 nm. (A), (B) The variations of $D_2$ and $D_{int}$ of antisymmetric modes with the gap. (C), (D) The variations of the $D_2$ and $D_{int}$ of antisymmetric mode with $W_2$.

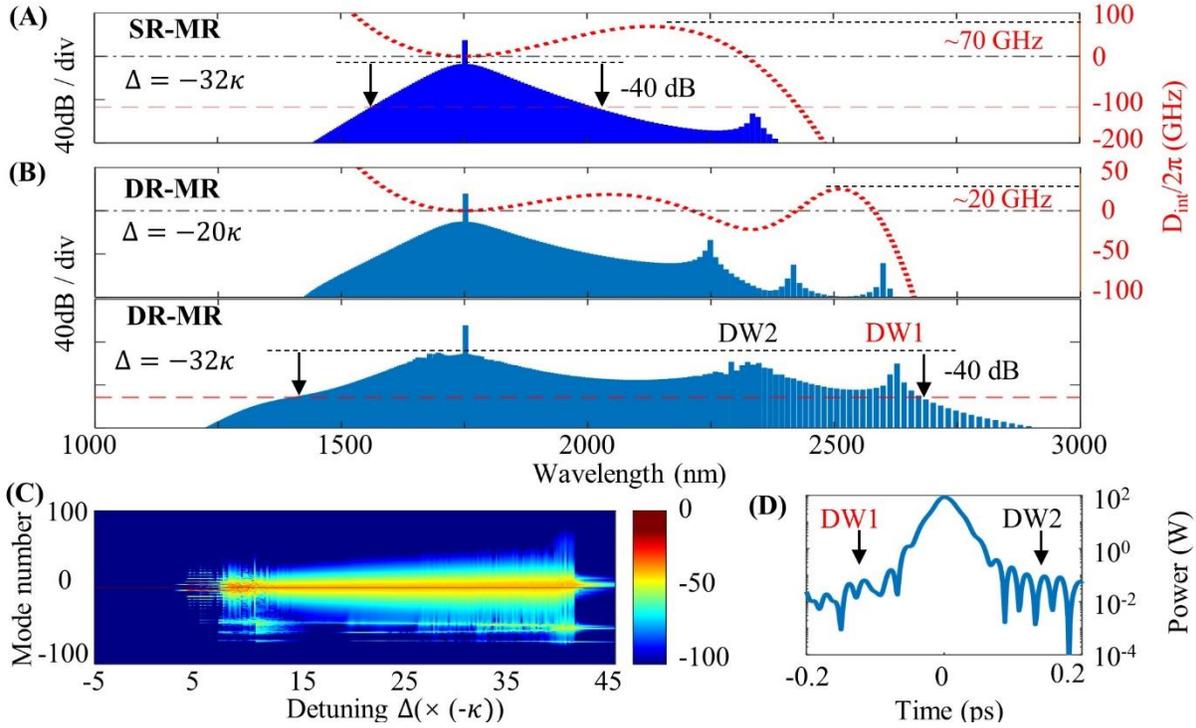

FIGURE 3. The generations of optical frequency combs in different microresonators. (A) The integrated dispersion (right axis) and frequency comb spectrum (left axis) of the SRM at the detuning $\Delta = -32\kappa$ with a -40 dB bandwidth of only 449.89 nm (1564.84 nm - 2014.73 nm). (B) Upper panel: the integrated dispersion of DRM with an ultra-flat and small dispersion configuration (right axis). The corresponding frequency comb spectrum from LLE simulation at detuning $\Delta = -32\kappa$ (left axis) features multiple DWs at phase-matching wavelengths. Lower panel: octave comb spectrum ranging from 1227.25 nm to 2912.87 nm at the same detuning $\Delta = -32\kappa$ as the spectrum in (A), with a -40 dB bandwidth of 1265.84 nm (1405.63 nm - 2671.47 nm). (C) Spectral evolution with detuning in the DRM. The pump frequency detunes from $5\kappa$ to $-45\kappa$. (D) The temporal waveform at $\Delta = -32\kappa$, indicates two typical DWs tails (labeled as DW1 and DW2) along with a soliton pulse.

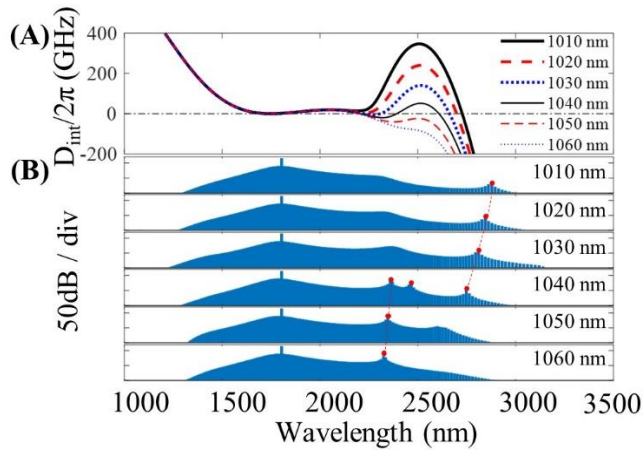

FIGURE 4. The spectral location tailoring of DWs by changing the outer microresonator width $W_2$ in the DRMs. (A) Integrated dispersion curves with $W_2$ varying from 1010 nm to 1060 nm. The other simulated parameters are fixed at R = 30 μm, $W_1$ = 2200 nm, H = 650 nm, gap = 800 nm. (B) The corresponding output spectra with different $W_2$, which allows spectral shaping of frequency combs and enhancement of the comb power at the desired waveband. The pump detuning is fixed at $-32\kappa$ for comparison.

Table 1. simulated geometric parameters of SRM and DRM

| Parameters | SRM | DRM |
|---|---|---|
| Pump frequency $v_p$ (THz) | 170.92 | 171.31 |
| Nonlinear refractive index $n_2$ (m$^2$/W) | $1.4 \times 10^{-18}$ | $1.4 \times 10^{-18}$ |
| Intrinsic quality factor $Q_i$ | $2 \times 10^6$ | $2 \times 10^6$ |
| External coupling factor $Q_c$ | $2 \times 10^6$ | $2 \times 10^6$ |
| Ring radius (μm) | 30 | $R_{in}$ = 30 |
| Effective index $A_{eff}$ (μm$^2$) | 1.1600 | 1.4156 |
| Input pump power $P_{in}$ (mW) | 40 | 40 |
| Free spectral range FSR (GHz) | 618.57 | 618.56 |
| Cross-section geometry (nm) (width $\times$ height, gap) | $1600 \times 650$ | Inner = $2200 \times 650$ Outer = $1050 \times 650$ gap = 825 |